\documentclass[sigconf]{acmart} 
\AtBeginDocument{%
  }

\usepackage{tikz}
\usetikzlibrary{arrows.meta,positioning,fit,calc,shapes,backgrounds}
\usepackage{graphicx}
\usepackage{subcaption}
\usepackage{booktabs}
\usepackage{adjustbox}
\usepackage{tabularx}
\renewcommand{\tabularxcolumn}[1]{m{#1}}

\usepackage[table,dvipsnames,svgnames]{xcolor}
\usepackage{multirow}
\usepackage{threeparttable}
\usepackage{placeins}

\definecolor{lighterpink}{RGB}{255, 242, 230}
\definecolor{lighterblue}{RGB}{230, 242, 255}

\newcommand{\customsize}{\footnotesize}
\newcolumntype{C}{>{\centering\arraybackslash}X} 
\newenvironment{smalltabularx}{\customsize \tabularx}{\endtabularx}

\newcommand{\sys}{\textsc{MEvince}}

\acmYear{2027}
\copyrightyear{2027}
\acmConference[WWW '27]{The ACM Web Conference 2027}{May 10--14, 2027}{Dublin, Ireland}

\begin{document}

\title{\sys{}: Knowledge-Constrained LLM Reasoning for Technique Identification from Malware Reports} 


\author{Hongyu She}
\affiliation{%
  \institution{Zhejiang University}
  \city{Hangzhou}
  \country{China}
}
\email{hongyushe@zju.edu.cn}

\author{Yiling He}
\correspondingauthor
\affiliation{%
  \institution{University College London}
  \city{London}
  \country{United Kingdom}
  }
\email{heyilinge0@gmail.com}
\authornote{Corresponding Author.}

\author{Mengda Xie}
\affiliation{%
  \institution{Jiangsu Police Institute}
  \city{Nanjing}
  \country{China}
  }
\email{985982441@qq.com}

\author{Yiping Liu}
\affiliation{%
  \institution{Alibaba Group}
  \city{Beijing}
  \country{China}
}
\email{ypinglinkle@gmail.com}

\author{Xingzhi Qian}
\affiliation{%
  \institution{University College London}
  \city{London}
  \country{United Kingdom}
  }
\email{xingzhi.qian.23@ucl.ac.uk}

\author{Zhan Qin}
\affiliation{%
  \institution{Zhejiang University}
  \city{Hangzhou}
  \country{China}
}
\email{qinzhan@zju.edu.cn}

\author{Lorenzo Cavallaro}
\affiliation{%
  \institution{University College London}
  \city{London}
  \country{United Kingdom}
  }
\email{l.cavallaro@ucl.ac.uk}


\begin{abstract}
Mapping behavioral intelligence in malware analysis reports to ATT\&CK techniques supports threat analysis and defense, but reliable identification remains challenging.
Supporting evidence is often scattered across long, heterogeneous reports, while closely related techniques share behavioral cues and differ only in subtle but decisive details.
While large language models~(LLMs) can use external knowledge to interpret reports, they may assign techniques based on plausible associations rather than sufficient evidence.

We present \sys{}, a knowledge-constrained LLM reasoning framework for report-level malware technique identification. 
Its core is a shared behavioral representation connecting report observations to technique definitions, specifying what evidence to extract and how it supports technique mapping. 
This representation is built through agentic mining of technique documentation and supports two online stages:
1)~\textit{primitive-guided extraction} uses behavioral primitives to structure LLM-extracted observations, link them to source passages, and group dispersed evidence;
2)~\textit{evidence-aligned retrieval} matches this evidence to technique profiles through the same concepts, retaining likely candidates and confusable alternatives for LLM comparison of distinguishing behaviors.
Controls at both stages validate evidence references and check technique-specific requirements. 
The representation also supports adaptation without parameter updates, where offline agentic audits diagnose concept weaknesses and refine extraction or retrieval policies.

We evaluate \sys{} against $11$ baselines spanning four method families using a dataset of $610$ public malware reports referenced by MITRE ATT\&CK. 
\sys{} improves exact-ID F1 over the strongest baseline by $17.4\%$, and its concept-guided retrieval increases gold-label candidate coverage from $31.6\%$ to $47.9\%$. With a small adaptation set, offline refinement yields further gains and outperforms baselines that update model parameters.
\end{abstract}




\maketitle

\section{Introduction}
Security vendors and incident-response teams routinely publish cyber threat intelligence (CTI) reports on the Web, documenting how malware and threat actors operate in real incidents~\cite{furumoto2025survey}. 
Mapping these reports to MITRE ATT\&CK techniques makes the reported observations searchable and comparable, supporting detection engineering, incident triage, and knowledge-base maintenance~\cite{mitre_attack,sharma2023ttp}. 
However, reliably automating this mapping while preserving evidence provenance remains challenging.
On the report side, Web reports are long and heterogeneous, mix observed behavior with background and mitigation text, and often distribute the evidence for a single technique across multiple sentences or sections.
Meanwhile, on the technique side, closely related ATT\&CK techniques often share behavioral cues and differ in the fine-grained details required to distinguish them.

Prior work generally falls into four categories: classification~\cite{rani2024ttpxhunter}, supervised fine-tuning (SFT)~\cite{chen2025aecr}, NER-style extraction~\cite{husari2017ttpdrill}, and context-driven LLM reasoning~\cite{guru2025technique,fayyazi2024advancing}. Classifiers and SFT models can learn recurring label patterns, but typically require task-specific annotations and a predefined label set. NER-style systems preserve explicit mentions but struggle with implicit behaviors~\cite{buchel2025sok} or evidence that must be composed across a report. Recent context-driven LLM methods, such as direct prompting~\cite{guru2025technique} and retrieval-augmented generation (RAG)~\cite{fayyazi2024advancing}, reduce the need for task-specific training. However, retrieved technique descriptions provide background knowledge rather than evidence that a behavior occurred in the report. 

As our motivating analysis of a real-world malware report illustrates~(\autoref{tab:case_blada_selective}), direct prompting misses distributed behavioral clues, whereas RAG recovers additional parent-level techniques but also produces substantially more false positives. These contrasting failure modes stem from two underlying gaps.
First, representation mismatch separates concrete report operations from abstract ATT\&CK mechanisms. Without domain knowledge specifying which behavioral primitives to preserve, extraction loses the context needed to compose dispersed clues, and text-based retrieval matches surface keywords without distinguishing closely related techniques.
Second, unregulated evidence authority blurs the distinct roles of observed report spans as source evidence, technique profiles as interpretive knowledge, and retrieved candidates as unverified hypotheses, leaving models to mistake retrieval rank or plausible explanations for factual support.
Together, these two gaps cause existing methods to either miss fine-grained techniques or produce unsupported overpredictions that burden analysts, motivating knowledge-constrained reasoning to govern what behavioral evidence is extracted, compared, and accepted.


We present \sys{}, a knowledge-constrained LLM reasoning framework for report-level malware technique identification. The key principle is that external knowledge guides interpretation, while every assignment must be supported by observed report evidence. At the core of \sys{} is a shared concept foundation built through offline agentic mining of technique description documentation, which connects report observations to technique profiles in terms of actions, objects, goals, and mechanisms. This representation supports two online stages. \textit{Primitive-guided extraction} structures LLM-extracted observations, links them to source passages, and groups dispersed clues. \textit{Evidence-aligned retrieval} then matches this structured evidence against technique profiles while retaining likely candidates and confusable alternatives for LLM comparison of distinguishing behaviors. To reduce hallucination, deterministic controls at both stages validate source references and enforce technique-specific evidence requirements before acceptance. To additionally support light-weight adaptation, an evidence memory module is designed to learn from labeled examples and refines candidate selection without fine-tuning model parameters. 


We evaluate \sys{} on public malware reports referenced by MITRE ATT\&CK. \sys{} achieves macro F1 scores of 0.310 at exact-ID granularity and 0.386 at parent-technique granularity, outperforming the baselines under the same evaluation setting. Controlled ablations show that concept-guided retrieval improves candidate coverage beyond evidence selection alone, while claim verification reduces false-positive assignments. Evidence-memory revision yields further gains without updating model parameters.

This work makes three contributions:
\begin{itemize}
    \item We introduce a shared behavioral representation that connects source-linked report observations to ATT\&CK technique definitions through behavioral primitives, while explicitly distinguishing observed report evidence from external knowledge and unverified hypotheses.
    
    \item We develop a knowledge-constrained reasoning pipeline that combines primitive-guided extraction and evidence-aligned retrieval under two-stage verification controls, alongside a parameter-free evidence memory that refines candidate selection.

    \item We systematically evaluate \sys{} on $610$ malware reports against $11$ baselines across four method families, using controlled ablations and provenance audits to demonstrate a $17.4\%$ exact-ID F1 gain and verifiable source traceability.
\end{itemize}

\begin{table*}[t]
\centering
\caption{Technique mappings for the njRAT report.}
\label{tab:case_blada_selective}
\footnotesize
\setlength{\tabcolsep}{4pt}
\renewcommand{\arraystretch}{1.12}
\begin{tabularx}{\linewidth}{lXXX}
\toprule
\textbf{Method} & \textbf{Representative Correct Mappings} & \textbf{Representative Errors (FNs / FPs)} & \textbf{Observation} \\
\midrule
\textbf{Direct}
& Replication Through Removable Media (\texttt{T1091}); PowerShell (\texttt{T1059.001}); Web Protocols (\texttt{T1071.001}); Registry Run Keys / Startup Folder (\texttt{T1547.001})
& \textcolor{blue}{16 FNs}, including Query Registry (\texttt{T1012}), Process Discovery (\texttt{T1057}), Modify Registry (\texttt{T1112}), and Non-Standard Port (\texttt{T1571}); \textcolor{red}{3 FPs}
& Captures five salient local behaviors, but misses when the mapping depends on composing weaker evidence distributed across the report. \\ \midrule

\textbf{CTIKG}
& [Sub-technique Level]\newline\textcolor{DarkGreen}{Adds}: File Deletion (\texttt{T1070.004})
& \textcolor{blue}{15 FNs (1 fewer)}; \textcolor{red}{6 FPs (3 more)}, including Spearphishing Attachment (\texttt{T1566.001}), Archive via Utility (\texttt{T1560.001}), and Local Data Staging (\texttt{T1074.001})
& Preserves structured intermediates with local relations, while higher pipeline complexity does not clearly improve final mapping quality. \\ \midrule

\textbf{RAG}
& [Parent-technique Level]\newline\textcolor{DarkGreen}{Adds}: Ingress Tool Transfer (\texttt{T1105}); Dynamic Resolution (\texttt{T1568})
& \textcolor{blue}{18 FNs (2 more)}; \textcolor{red}{12 FPs (9 more)}, including AutoHotKey \& AutoIT (\texttt{T1059.010}), Remote Access Tools (\texttt{T1219}), Active Setup (\texttt{T1547.014}), and Hide Infrastructure (\texttt{T1665})
& Recovers additional true parent-level positives, but also raises substantially more FPs, suggesting that the model tends to overcommit to retrieved candidates even when evidence is weak. \\
\bottomrule
\end{tabularx}
\end{table*}

\section{Background and Related Work}

This section reviews existing work on report-level technique identification, with additional related work discussed in Appendix~\ref{sec:additional_related_work}.

\subsection{CTI for TTP Identification}

Cyber Threat Intelligence (CTI) reports describe adversarial behaviors, contextual information, and analyst observations in natural language, providing a rich but unstructured knowledge base for malware analysis~\cite{sharma2023ttp,xuan2026identifying,guo2025intelliradar}.
A fundamental task in turning this information into actionable insights for systematic analysis is to map the reported behaviors to the MITRE ATT\&CK framework~\cite{mitre_attack}, which organizes adversarial tactics, techniques and procedures (TTPs) into a structured knowledge base.
Building on prior taxonomies~\cite{buchel2025sok}, we organize existing methods into four representative paradigms, extending the categorization to include recent reasoning-oriented approaches enabled by large language models (LLMs).

\noindent\textbf{Classification-based Methods.}
A common approach formulates TTP identification as multi-label classification over a label set defined by a specific ATT\&CK version.\footnote{The label set varies across versions: \url{https://attack.mitre.org/resources/versions}} These methods take entire reports~\cite{liu2022threat,legoy2020rcatt}, or more commonly, individual sentences~\cite{rani2024ttpxhunter,rahman2024alert} as input and require annotations at the corresponding granularity. 
Earlier approaches use features such as TF--IDF with conventional classifiers~\cite{ayoade2018automated}, whereas recent systems use BERT~\cite{devlin2019bert} and domain-adapted encoders to obtain contextual representations for downstream classification~\cite{jin2023darkbert,ranade2021cybert,bayer2024cysecbert}.


\noindent\textbf{NER-style Methods.}
Another line of work uses Named Entity Recognition (NER)~\cite{li2022nersurvey} and rule-based NLP processing~\cite{li2022attackg}. 
These methods extract threat actions and indicators and map them through ontologies, lexical rules, or patterns ~\cite{husari2017ttpdrill,husari2018actionminer,zhang2021ex}. 
AttacKG~\cite{li2022attackg} further models dependencies among extracted entities and uses graph matching for TTP mapping. By retaining intermediate evidence as explicit textual or structured representations, these approaches make the mapping process more interpretable.

\noindent\textbf{SFT-based LLM Methods.}
Supervised fine-tuning (SFT) adapts generative LLMs to produce TTP predictions as text. These methods typically convert sentence-level CTI annotations into instruction-style training examples, pairing an input and task instruction with the corresponding TTP labels~\cite{chen2025aecr,fieblinger2024actionable}. 
Common backbones include Llama and Qwen~\cite{grattafiori2024llama,yang2024qwen2}. Compared with classification-based methods, SFT-based approaches support more flexible output formats, but the mapping usually remains end-to-end.

\noindent\textbf{Context-driven LLM Methods.}
Context-driven methods adapt LLM behavior by refining the information supplied at inference time. Direct prompting uses security ontology-aligned instructions, sometimes supplemented with few-shot examples~\cite{cheng2024ctinexus}. 
Retrieval-augmented generation (RAG)~\cite{lewis2020retrieval} further incorporates external ATT\&CK knowledge to ground TTP mapping~\cite{guru2025technique,fayyazi2024advancing}.
Multi-stage systems such as CTIKG~\cite{huang2024ctikg} segment long CTI reports and employ multiple LLM agents with dual-memory mechanisms to extract and merge entity relations. These designs reduce reliance on task-specific training and can retain intermediate clues.

\subsection{Why Context-driven LLMs}

The recent SoK on CTI-based TTP identification~\cite{buchel2025sok} finds that generative LLMs are not uniformly superior: SFT is broadly comparable to classification when labeled data is available, while training-free prompting only slightly outperforms NER baselines. 
Beyond benchmark accuracy, CTI-to-TTP mapping also exposes two practical tensions that motivate context-driven methods.

\noindent\textbf{Supervision reliance limits adaptability.}
Classification and SFT-based methods depend on task-specific supervision over a fixed label space. As ATT\&CK releases introduce or revise techniques, these methods face recurring maintenance to accommodate changes in the target ontology~\cite{he2025combating}.
Labeled CTI data are also limited: TRAM2 covers 50 techniques in 4,070 labeled sentences from 150 reports~\cite{ross2023tram2}. 
Prior work further reports challenges in open-set settings and performance degradation when moving from sentence-level to document-level training~\cite{buchel2025sok}. 
These issues make repeated retraining difficult to sustain and motivate methods that can adapt primarily at inference time.

\noindent\textbf{Evidence preservation supports analysis and refinement.}
The interpretation of a local cue depends on its surrounding context, and reliable mapping may require combining multiple observations. 
For instance, a domain, an HTTP request, and periodic beaconing may support command-and-control only when considered together. 
End-to-end models obscure this composition, making incorrect mappings difficult to diagnose or refine, while NER pipelines preserve limited local structure and struggle with implicit references
~\cite{chen2023learning,perera2015implicit}. 
Context-driven LLMs provide a way to reason over explicit intermediate evidence, provided that the pipeline preserves and verifies it. 
In \sys{}, each accepted technique retains its source spans and matched concepts, making both the decision and its false-positive controls inspectable. 


\section{From Bottlenecks to Design Insights}
Despite the promise, current context-driven LLM approaches exhibit notable limitations in CTI-to-TTP mapping. In this section, we first analyze their bottlenecks through a case study, and then derive corresponding design insights.

\subsection{Bottleneck Analysis}

Our case study is based on a Trend Micro report on the njRAT worm~\cite{trendmicro2018njrat}, which describes multiple behaviors that cover 21 techniques and sub-techniques, making it useful for analyzing how current methods handle semantically dense CTI evidence. 
We present the results in Table~\ref{tab:case_blada_selective} and discuss three main observations below.

\noindent\textbf{Direct prompting remains under-constrained.}
Direct prompting identifies salient behaviors supported by explicit local cues, such as \textit{Replication Through Removable Media} (\texttt{T1091}) directly stated as ``installs a hidden copy of itself on any removable drive found on the infected system'', but misses labels that require contextual interpretation or the combination of weaker clues. 
For example, the report states that the backdoor communicates with a DuckDNS domain over port 1177. Mapping this observation to \textit{Non-Standard Port} (\texttt{T1571}) requires treating the port as part of C2 communication rather than as an isolated network detail.

\noindent\textbf{Complex extraction does not directly solve mapping.}
CTIKG uses multiple LLM stages to extract and merge structured knowledge from long reports~\cite{huang2024ctikg}.
Compared with Direct Prompt, it yields a net gain of one correct mapping but three additional false positives.
For example, it infers \textit{Spearphishing Attachment} (\texttt{T1566.001}) from a statement about a malicious file arriving in the system, although the report explicitly states that its arrival mechanism is unknown.
It also infers \textit{Archive via Utility} (\texttt{T1560.001}) from a statement about sending stolen information to the C2 server, without evidence of archiving.
These errors show that structured extraction alone does not ensure sufficient evidence for a technique assignment.

\noindent\textbf{RAG supplies ontology knowledge but not decision evidence.}
RAG is a natural way to supply ATT\&CK knowledge at inference time~\cite{guru2025technique,fayyazi2024advancing}, but retrieved knowledge does not ensure a better decision.
In this case, RAG reduces parent-level false negatives (12 versus 14 for Direct) but increases exact-ID false negatives (18 versus 16), recovering some families without resolving their sub-techniques.
RAG also produces 12 exact-ID false positives, compared with three for Direct Prompt.
This suggests that retrieval can supply relevant ontology knowledge, but does not decide whether the report contains sufficient evidence.


\subsection{Design Insights}

While extraction can isolate threat-relevant clues and RAG can provide external technique knowledge, the case study suggests that their current forms do not yet translate into robust gains.
Our insights are to organize these components around the evidence required for a technique assignment, which are described as follows. 

\noindent\textbf{Insight 1: Extraction should preserve TTP-relevant primitives.}
Current extraction pipelines often optimize for generic entities, relations, or graph completeness.
However, TTP mapping depends on more specific evidence primitives, such as an action, its target artifact, and contextual qualifiers.
Extraction should therefore be designed not merely to compress the report into triples, but to preserve the behavioral evidence needed to distinguish candidate techniques and link it to the original passages.

\noindent\textbf{Insight 2: RAG should retrieve over evidence-aligned representations.}
RAG is useful because the knowledge base provides rich technique descriptions, but retrieval quality depends heavily on the representation used for matching.
Instead of retrieving candidates from raw report text alone, retrieval should be driven by the behavioral primitives extracted from the report.
Representing observed behavior and candidate techniques in the same terms makes their comparison more targeted and helps distinguish techniques that share broad keywords.

\noindent\textbf{Insight 3: LLM outputs require explicit false-positive control.}
Grounding alone is not sufficient.
Our case study shows that even when relevant technique knowledge is provided, the LLM can still overcommit to weakly supported labels, leading to false positives.
The final mapping stage should therefore verify whether the report provides the evidence required for each proposed technique, and similarly, generated observations should be checked against their source passages during extraction.
These checks allow external knowledge to guide interpretation while keeping technique assignments grounded in the report.







\section{\sys{}}
\subsection{System Overview}

\begin{figure}[t]
    \centering
    \includegraphics[width=\linewidth]{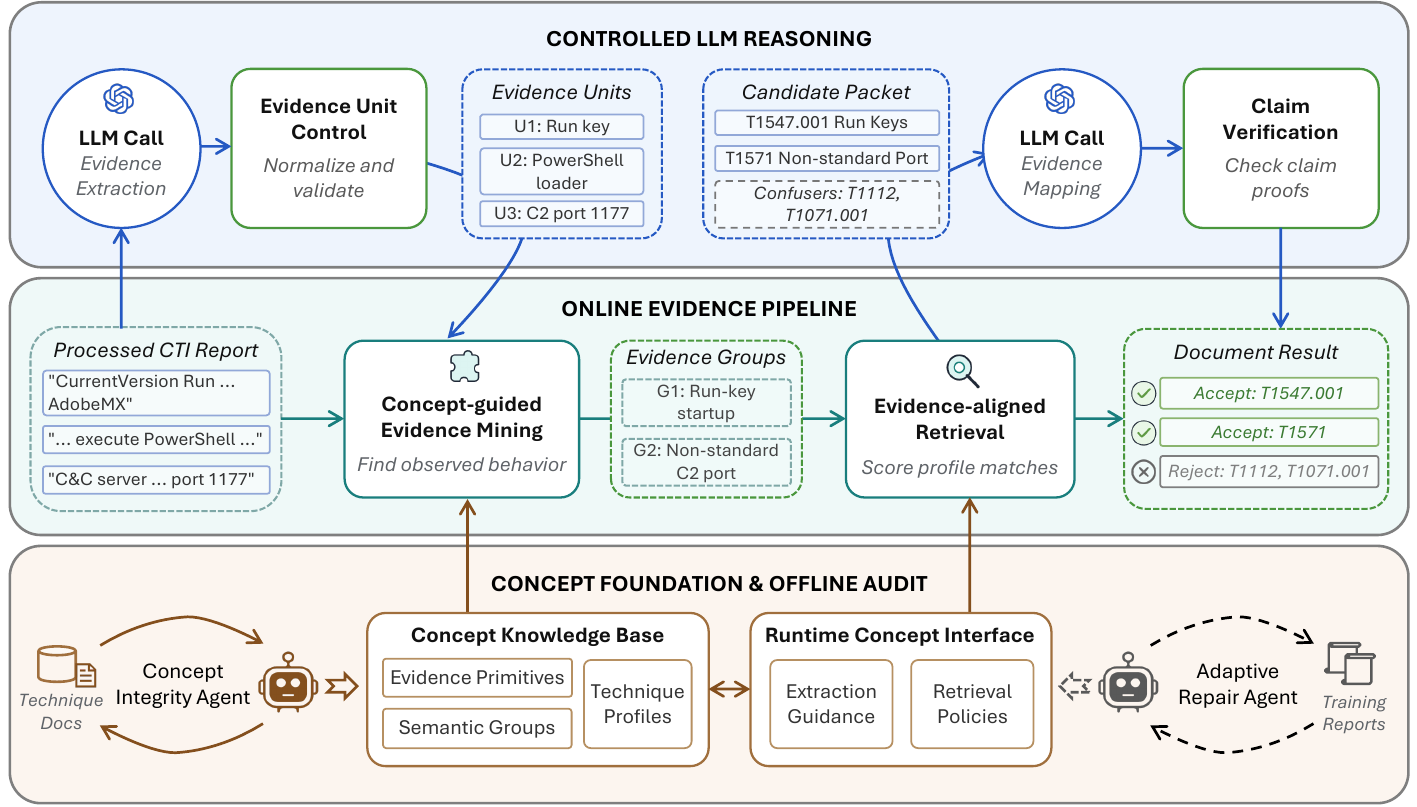}
    \Description{MEvince has three layers. Controlled LLM reasoning extracts evidence and maps candidates under deterministic checks. The online evidence pipeline mines concepts, forms evidence groups, retrieves candidates, and produces document decisions. A concept foundation and offline audit supply the knowledge used by both online modules.}
    \caption{\sys{} system architecture.}
    \label{fig:architecture}
\end{figure}

Figure~\ref{fig:architecture} presents the workflow of \sys{}. Given a malware report, the system extracts and organizes behavioral evidence, retrieves candidate techniques for LLM assessment, and verifies the resulting claims before producing report-level technique assignments.
At the core of this workflow is a concept foundation built offline from technique documentation (Section~\ref{sec:concept_foundation}). It connects report observations with technique knowledge through a shared behavioral representation, guiding two online modules. \textit{Primitive-guided extraction} structures behavioral observations, preserves their source spans, and groups related evidence (Section~\ref{sec:evidence_extraction}). \textit{Evidence-aligned retrieval} uses the same concepts to retrieve likely techniques and confusable alternatives for LLM comparison (Section~\ref{sec:inference}). Deterministic controls check evidence references after extraction and technique-specific requirements after mapping (Section~\ref{sec:output_control}).
An optional adaptive block uses an evidence memory learned from labeled reports to revise candidate priority (Section~\ref{sec:evidence_memory}).

\subsection{Concept Foundation and Knowledge Base}
\label{sec:concept_foundation}

\sys{} connects reports and technique definitions through behavioral primitives that provide a common basis on which evidence mining and retrieval can be built.

\noindent\textbf{Behavioral primitives.}
We define \emph{anchors} as consistent concept labels mined from reports that can identify behavior or an artifact in context.
The vocabulary is the disjoint union $\mathcal{C}=\mathcal{C}_A\mathbin{\dot\cup}\mathcal{C}_O\mathbin{\dot\cup}\mathcal{C}_G\mathbin{\dot\cup}\mathcal{C}_M$ of action, object, goal, and mechanism concepts. Actions and objects identify an operation and its target; goals describe its operational purpose; mechanisms capture how it is carried out. The latter two dimensions help distinguish techniques involving similar operations. Table~\ref{tab:concept_examples} in Appendix~\ref{sec:concept_examples} summarizes the four dimensions, their inventory sizes, and representative entries. 
For an evidence item $x$, we write $\Phi(x)=(A_x,O_x,G_x,M_x)$, where each component contains labels from its corresponding vocabulary. 
These dimensions provide a reference frame for characterizing behavioral evidence. 
For example, modifying a registry Run key may yield the action \emph{modify}, the object \emph{registry key}, the goal \emph{persistence}, and the mechanism \emph{registry-run-key startup}. 

\noindent\textbf{Knowledge organization and runtime interface.}
The resulting knowledge base is
\begin{equation}
    \mathcal{K}=\bigl(\mathcal{C},\mathcal{G},\{P_t\}_{t\in\mathcal{T}}\bigr),
    \label{eq:concept_foundation}
\end{equation}
where $\mathcal{G}$ contains reusable semantic groups, $P_t$ represents a technique profile, and $\mathcal{T}$ indexes techniques and sub-techniques. 
Semantic groups capture recurring behavioral roles and combinations of primitives, such as startup persistence or payload delivery followed by execution. During online evidence mining, \sys{} uses these groups to organize related report observations into \emph{evidence groups}, which retain their members’ source references and evidential status.
Each technique $t$ has a profile $P_t=(\Phi_t,S_t,R_t,Q_t)$, where $\Phi_t$ contains its concept sets, $S_t$ its salient phrases and aliases, $R_t$ its ontology neighbors and confusers (i.e., plausible alternative techniques that differ in their defining behavior), and $Q_t$ its evidence requirements (i.e., artifacts, relations, or mechanisms needed to distinguish $t$ from the alternatives). 

The Runtime Concept Interface uses this knowledge base to provide extraction guidance and retrieval policies, which will be expanded in the subsequent sections. The extraction schema specifies which behavioral details the LLM should preserve. The deterministic miner then uses the vocabulary and anchors to map the extracted descriptions to concept labels. Retrieval policies rank technique profiles using $\Phi_t$ and $S_t$ and retain relevant alternatives using $R_t$. The requirements in $Q_t$ govern acceptance during verification. 

\noindent\textbf{Offline concept mining.}
To construct such a foundation, we develop a Concept Integrity Agent that uses a staged agent-script workflow. 
First, the agent proposes canonical primitives and their anchors. Deterministic scripts then identify unresolved phrases, overly broad labels, and missing distinctions to guide vocabulary revisions. Second, the agent composes primitives into reusable semantic groups and annotates technique profiles in the same representation. We check whether related profiles overlap too much or omit mechanisms and conditions that distinguish the techniques. Third, we test whether the miner can recognize the required concepts in report evidence and whether retrieval returns the intended techniques and plausible alternatives. If a check fails, we revise the relevant concepts or runtime rules and repeat the check.

\subsection{Primitive-Guided Evidence Extraction and Mining}
\label{sec:evidence_extraction}

For each report segment $s$, this module extracts evidence units and organizes related observations into local groups in the shared concept space. 
The process is summarized as
\begin{equation}
    Z_s=M_{\mathcal K}\!\left(
    V_{\mathrm{ext}}\!\left(E_\theta(s;\mathcal D_{\mathcal C}),s\right)
    \right),
    \label{eq:evidence_mining}
\end{equation}
where $\mathcal D_{\mathcal C}$ is the extraction schema based on the primitive dimensions in Section~\ref{sec:concept_foundation}, $E_\theta$ is the extraction LLM, $V_{\mathrm{ext}}$ performs evidence-unit control, and $M_{\mathcal K}$ is the deterministic miner that canonicalizes observations and composes local evidence using $\mathcal K$.

For pre-processing, \sys{} partitions a report along its document structure, retaining headings, segment types, source identifiers, and offsets. Adjacent paragraphs under the same heading are combined when local context is needed. Empty fragments and indicator-dominated fragments that contain no behavioral information are skipped. 
Then there are three main steps.

\noindent\textbf{Step 1: preserve behavioral evidence.} 
The extraction schema defines fields for actor, action, object, goal, and mechanism to guide the LLM in preserving the behavioral information needed for subsequent mapping. For each segment, the model produces an evidence ledger $L_s=(U_s,V_s)$. Each unit $u\in U_s$ contains a minimal, complete behavioral statement, its source spans, typed fields, mapping relevance, and an evidential status. The statuses distinguish behavior reported as having occurred (\emph{observed}), background information (\emph{contextual}), defensive advice (\emph{advisory}), descriptions of malware functions (\emph{capability}), unclear or weakly supported statements (\emph{ambiguous}), and speculation (\emph{hypothetical}). A literal $v\in V_s$ preserves the exact text of an artifact, such as a command, registry path, filename, URL, or domain. It can also carry a short note describing how that artifact relates to the surrounding evidence.

The model keeps distinct actions as separate units and leaves unsupported fields empty. Modal statements and capability lists retain their qualifiers rather than being rewritten as executed behavior. 

\noindent\textbf{Step 2: align to the source and canonicalize.} Evidence-unit control first normalizes and source-aligns the
extracted ledger. 
For each unit $u$, the concept miner maps its extracted fields and source text to $\Phi(u)$ using the vocabulary and anchors in $\mathcal K$. Conservative anchor requirements limit generic matches involving words such as \emph{access}, \emph{create}, or \emph{service}. High-value literals remain attached because a short command or path often supplies the mechanism needed to distinguish related techniques.


\noindent\textbf{Step 3: compose local evidence.} 
When behavioral evidence spans multiple statements, the miner groups these units and literals within a local mapping context using the reusable behavioral patterns. For example, a Run key, its value, and the referenced executable may jointly describe startup behavior. For a group $g$, $\Phi(g)$ combines the concepts of its members. The miner adds information about how the behavior is carried out only when the matched anchors and relations support it.
The group retains the identifiers, source spans, and evidential statuses of its members. A group can be treated as observed evidence only if at least one member reports that the behavior occurred. A group containing only literals from background or advisory text remains contextual.

The same pass may record a weak hypothesis when a relation is plausible but incomplete. For example, background mentions of phishing may trigger a phishing-family check while recording that evidence of an observed lure, attachment, or link is still missing. 
Such hypotheses will be passed to guide retrieval.

\subsection{Evidence-Aligned Retrieval and Technique Mapping}
\label{sec:inference}

This module uses the shared concept representation to connect extracted observations with candidate techniques. 

\noindent\textbf{Step 1: retrieve compatible profiles.} Retrieval targets are constructed from the evidence units, local groups, and marked hypotheses produced in Section~\ref{sec:evidence_extraction}. For each target $z\in Z_s$, candidate generation is defined as
\begin{equation}
    \mathcal Q_z=\operatorname{Expand}\!\left(
    \operatorname{TopK}_{t\in\mathcal T}
    \operatorname{Match}(\Phi(z),P_t)\right),
    \label{eq:concept_retrieval}
\end{equation}
where $\operatorname{Match}$ ranks technique profiles by concept compatibility, $\operatorname{TopK}$ selects up to $K$ highest-ranked profiles per target as initial candidates, and $\operatorname{Expand}$ retains additional techniques indicated by behavioral rules and uses $R_t$ to add relevant parents, siblings, and known confusers. For marked hypotheses, $\Phi(z)$ encodes concepts that guide retrieval without constituting observed evidence.

Matching compares $\Phi(z)$ with the profile concepts $\Phi_t$, using salient phrases and aliases in $S_t$ as additional signals. Specific mechanism matches receive greater priority than broad lexical overlap, with goals, actions, and objects providing complementary support. 
Evidential status also affects ranking, giving contextual material less weight than observed behavior.

\noindent\textbf{Step 2: construct the comparison context.} Entries for the same technique are merged under its canonical identifier. The merged entry records which concepts matched, which evidence is linked to it, how it was retrieved, and what evidence is missing.
An added parent profile without a structural match remains an alternative and does not inherit its child's support.
The resulting candidate packet allows the mapper to compare retrieved techniques and nearby alternatives against the same evidence. For example, a Run-key observation may retrieve both startup persistence and registry modification. The key's startup role and referenced executable provide evidence for distinguishing the behaviors. 

\noindent\textbf{Step 3: evaluate technique claims.} The mapping LLM receives a bounded packet containing source-linked evidence units and groups, candidate techniques, nearby alternatives, and acceptance conditions derived from $Q_t$. A selected subset of candidates forms the mandatory decision queue. For each candidate in the mandatory decision queue, it returns \emph{accept}, \emph{abstain}, or \emph{reject}, together with supporting evidence identifiers and structured proof gaps. The remaining candidates provide alternatives for comparison.
If none of the candidates fit, the model may propose another technique supported by observed report evidence. Such proposals must resolve to an indexed profile and satisfy the same evidence checks. 

\subsection{Output Control}
\label{sec:output_control}

\sys{} checks whether extracted observations can be traced back to the report and whether proposed technique assignments have sufficient support in the report. The two controls operate after extraction and mapping, respectively, so LLM responses are not treated as final decisions.

\noindent\textbf{Evidence-unit control.} An evidence unit records a behavioral statement from the report, its source spans, and its evidential status. In the Reaver example below, \texttt{U5} records XOR decryption of an embedded file as \emph{observed}: the report describes the behavior as having occurred. After extraction, the control normalizes these records, aligns quoted spans with the source segment, and records offsets for matched text. It assigns local identifiers to units and literals, and recovers selected artifacts, such as commands and registry paths, directly from the report. Subsequent stages use these identifiers within the corresponding report and segment to retrieve the original passages. 

\noindent\textbf{Claim verification.} A technique claim contains a candidate technique, a verdict, and evidence references. The verifier resolves these references and checks the claim against the report evidence and requirements $Q_t$. Acceptance requires observed behavior concerning the malware under analysis or explicit capability evidence permitted by a matching rule; background, advisory text, and hypotheses alone are insufficient. An \emph{accept} claim without valid citations becomes \emph{abstain}, where a \emph{proof gap} records missing or conflicting evidence and marks a \emph{blocking} reason. In the example below, the accepted T1140 claim cites \texttt{U5/G3}, whereas the mapper rejects T1057 with a blocking \texttt{no\_observed\_support} gap because the source segment contains no process-discovery behavior. 
\par\smallskip
\begingroup
\setlength{\fboxsep}{4pt}
\setlength{\fboxrule}{0.35pt}
\noindent\makebox[\linewidth][c]{%
\fcolorbox{black!35}{white}{%
\parbox{\dimexpr0.82\linewidth-2\fboxsep-2\fboxrule\relax}{%
\footnotesize
\setlength{\parindent}{0pt}
\textbf{Reaver: evidence and mapper decisions}\par\vspace{2pt}
\texttt{U5} (\emph{observed}): XOR decryption of an embedded file.\par\vspace{3pt}
\renewcommand{\arraystretch}{1.0}
\begin{tabular}{@{}>{\ttfamily}p{27pt}@{\hspace{6pt}}p{25pt}@{\hspace{6pt}}>{\raggedright\arraybackslash}p{\dimexpr\linewidth-64pt\relax}@{}}
T1140 & \emph{accept} & File decryption; cites \texttt{U5/G3}. \\
\addlinespace[2pt]
T1057 & \emph{reject} & No process discovery in this segment.\newline Gap: \texttt{no\_observed\_support} (\emph{blocking}).
\end{tabular}\par
}}}
\par
\endgroup
\smallskip

An \emph{abstain} claim may be accepted only after passing the same checks and a further assessment of its evidence support. The evidence must match the candidate profile through a mechanism, a salient phrase, or all three action--object--goal dimensions. The assessment considers whether the support is observed, how specifically it matches the candidate, retrieval strength, and the number of distinct supporting units and groups. Claims are then consolidated across segments, with supported sub-techniques taking precedence over redundant parents; each decision retains its verdict, claim and evidence references, proof gaps, and verification results.

\subsection{Offline Adaptation of the Retrieval Interface}
\label{sec:evidence_memory}

\sys{} optionally uses labeled training reports to adjust which retrieved techniques the LLM must explicitly judge. For each technique annotated in a training report, \sys{} retains up to two evidence units that describe observed behavior, are relevant to mapping, and align with that technique's profile. These examples form a positive evidence memory; missing annotations are not negatives because report labels may omit behaviors.

Memory can move an alternative already in the retrieved packet into the mandatory decision queue, replacing its weakest-scoring member while keeping the base concept vocabulary unchanged. 
For each candidate, we compare its evidence texts with stored examples of the same technique. The highest cosine similarity between their embeddings gives the candidate's score. Eligibility requires observed support or capability support permitted by retrieval rules, and the technique must have examples from a minimum number of training reports. Eligible alternatives are considered in descending score order and must reach the threshold and score no lower than the replaced candidate. 
Exchanges are capped per packet; queue size remains unchanged.
Taking Figure~\ref{fig:training_case} as an example, the Reaver report describes placing a shortcut in a startup path, and T1547.009 (Shortcut Modification), initially a confuser, enters the queue through similar training examples. 


\section{Evaluation} 
We evaluate \sys{} on overall technique identification (\textbf{RQ1}), evidence selection and concept-based retrieval (\textbf{RQ2}), output control (\textbf{RQ3}), optional evidence-memory adaptation (\textbf{RQ4}), and online LLM workload (\textbf{RQ5}).

\subsection{Experimental Setup}
\label{sec:eva_setup}

\noindent\textbf{Dataset.} Our benchmark comprises 610 public malware-analysis reports referenced by the MITRE ATT\&CK Enterprise software knowledge base~\cite{mitre_attack}, with 6,095 report--technique assignments. We reserve 555 reports for optional in-domain adaptation (5,592 assignments) and use the remaining 55 for development evaluation (503 assignments across 164 technique identifiers). All main comparisons, including the evidence-memory comparison, use the same 55 evaluation reports. The partitions are report-disjoint but not software-disjoint: reports in different partitions may describe the same software entry. Gold labels are derived from software--technique relationships citing report URLs. 

\begin{table*}[!t]
\centering
\caption{Technique identification performance.}
\label{tab:technique_identification}
\begin{threeparttable}
    \renewcommand{\customsize}{\fontsize{7}{8.6}\selectfont}
    \setlength{\tabcolsep}{3.5pt}
    \renewcommand{\arraystretch}{1.00}
    \setlength{\aboverulesep}{1.5pt}
    \setlength{\belowrulesep}{1.5pt}
    \begin{smalltabularx}{.94\linewidth}{c >{\centering\arraybackslash}p{9em} CCCC @{\hspace{8pt}} CCCC}
    \toprule[0.6pt]
    \multicolumn{2}{c}{\textbf{Method}} & \multicolumn{4}{c}{\textbf{Exact ATT\&CK IDs}} & \multicolumn{4}{c}{\textbf{Parent IDs}} \\
    \cmidrule(lr){1-2} \cmidrule(lr){3-6} \cmidrule(lr){7-10}
    \textbf{Type} & \textbf{Name} & \textbf{F1} & \textbf{Recall} & \textbf{Prec.} & \textbf{Jacc.} & \textbf{F1} & \textbf{Recall} & \textbf{Prec.} & \textbf{Jacc.} \\
    \midrule[0.35pt]
    \multirow{2}{*}{Classifier} & RoBERTa & 0.250 & 0.234 & 0.294 & 0.158 & 0.351 & 0.369 & 0.389 & 0.233 \\
     & CySecBERT & 0.140 & 0.214 & 0.111 & 0.079 & 0.295 & \underline{0.503} & 0.225 & 0.183 \\
    \midrule[0.3pt]
    \multirow{3}{*}{Extraction} & NER-base & 0.105 & 0.166 & 0.091 & 0.059 & 0.151 & 0.212 & 0.139 & 0.089 \\
     & NER-full & 0.144 & 0.268 & 0.108 & 0.082 & 0.214 & 0.347 & 0.172 & 0.127 \\
     & AttacKG & 0.038 & 0.069 & 0.028 & 0.020 & 0.099 & 0.180 & 0.072 & 0.055 \\
    \midrule[0.3pt]
    \multirow{3}{*}{SFT} & Llama-3.1-8B & 0.244 & 0.221 & 0.303 & 0.154 & 0.350 & 0.357 & 0.404 & 0.232 \\
     & Llama-3.1-70B & \underline{0.264} & 0.235 & \underline{0.329} & \underline{0.169} & \underline{0.379} & 0.379 & \underline{0.438} & \underline{0.256} \\
     & Qwen2.5-14B & 0.255 & 0.245 & 0.288 & 0.161 & 0.363 & 0.395 & 0.385 & 0.241 \\
    \midrule[0.3pt]
    \multirow{3}{*}{Inference} & Direct Prompt & 0.189 & 0.194 & 0.205 & 0.112 & 0.271 & 0.286 & 0.286 & 0.169 \\
     & RAG & 0.251 & \textbf{0.424} & 0.191 & 0.149 & 0.336 & \textbf{0.529} & 0.270 & 0.211 \\
     & CTIKG & 0.164 & 0.167 & 0.174 & 0.097 & 0.262 & 0.294 & 0.267 & 0.161 \\
    \midrule[0.3pt]
    \textbf{Ours} & \textbf{\sys{}} & \textbf{0.310} & \underline{0.302} & \textbf{0.384} & \textbf{0.200} & \textbf{0.386} & 0.383 & \textbf{0.463} & \textbf{0.260} \\
    \bottomrule[0.6pt]
    \end{smalltabularx}

\end{threeparttable}
\end{table*}

\noindent\textbf{Baselines.} We compare 11 baselines spanning four families: document classification with RoBERTa~\cite{liu2019roberta} and CySecBERT~\cite{bayer2024cysecbert}; mention extraction and structured mapping with NER variants and AttacKG~\cite{li2022attackg}; supervised fine-tuning of Llama and Qwen~\cite{grattafiori2024llama,qwen2025qwen25}; and context-driven LLM methods, including Direct Prompt, RAG, and CTIKG~\cite{huang2024ctikg,fayyazi2024advancing,guru2025technique}. The GPT-based baselines and \sys{} use GPT-4o. Methods retain their implementation-specific prompts and decoding schedules; all predictions are normalized by the same evaluator. 
More details are provided in Appendix~\ref{sec:baseline_details}.

\noindent\textbf{Evaluation protocol.} Our main comparison evaluates \sys{} against the 11 baselines (Table~\ref{tab:technique_identification}). We separately evaluate optional evidence-memory adaptation alongside baselines adapted on our training partition (Table~\ref{tab:trained_technique_identification}). Some baselines in the main comparison use models trained on data from their original implementations. The evidence memory is constructed from training reports only; its configuration is selected through training-only validation, and both are fixed before evaluation.

We report report-level macro precision, recall, F1, and Jaccard at exact-ID and parent-technique granularity, with macro F1 as the primary metric. Exact-ID evaluation preserves technique and sub-technique identifiers, whereas parent-level evaluation maps sub-techniques to their parent identifiers. Component analyses also report aggregate error counts and micro F1.

\noindent\textbf{Implementation.} The knowledge base contains 651 technique and sub-technique profiles from ATT\&CK Enterprise v19.2. For each evidence target, retrieval selects up to four initial profiles and supplements them with techniques indicated by behavioral rules and ontology-based alternatives. \sys{} uses GPT-4o with temperature 1.0 and a 2,048-token output limit. Evidence-memory configuration details are provided in Appendix~\ref{sec:reproducibility}.

\subsection{Overall Technique Identification (RQ1)}

Table~\ref{tab:technique_identification} shows that \sys{} achieves the highest macro F1 and Jaccard at both granularities. Its exact-ID F1 reaches 0.310, exceeding the strongest baseline, Llama-3.1-70B (0.264), by 17.4\%. At the parent level, its F1 is 0.386, compared with 0.379 for the strongest baseline. The advantage is more pronounced at the exact-ID level, where evaluation requires preserving distinctions between techniques and sub-techniques.

Among the context-driven LLM methods, RAG improves on Direct Prompt, suggesting the value of external technique knowledge. However, neither RAG nor CTIKG matches \sys{} despite using the same LLM backbone. These results suggest that access to external knowledge alone is insufficient for reliable technique identification; how that knowledge is connected to report observations and how proposed assignments are verified also matter. Against RAG, a paired report bootstrap~\cite{efron1979bootstrap} gives \sys{} an exact-ID macro-F1 advantage of 0.060, with a 95\% percentile interval of $[0.014,0.106]$.
The precision--recall pattern provides further insight. RAG achieves the highest exact-ID recall (0.424), but its precision is only 0.191. In contrast, \sys{} approximately doubles precision to 0.384 while retaining the second-highest recall (0.302). At the parent level, the same trade-off holds: \sys{} achieves 0.463 precision and 0.383 recall, compared with 0.270 and 0.529 for RAG. These results are consistent with the separation of retrieval and verification in \sys{}: retrieved techniques remain hypotheses until supported by report evidence. Further discussion of this tradeoff and the study's limitations is provided in Appendix~\ref{sec:supplementary_discussion}.

\subsection{Extraction and Retrieval Ablation (RQ2)}
\label{sec:retrieval_ablation}

To separate evidence selection from concept-based matching, we compare four retrieval variants using the same archived reports and 651 technique profiles, with candidate budgets of 10, 20, and 50 per report. Candidates are deduplicated and ranked at the report level before applying each budget. Gold-label coverage measures the fraction of reference report--technique associations retrieved across reports.

\begin{figure}[t]
    \centering
    \includegraphics[width=\linewidth]{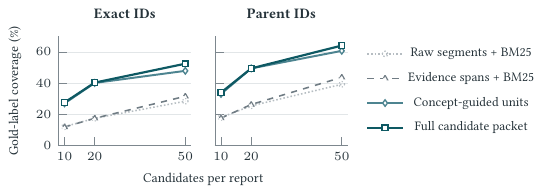}
    \Description{Two line charts compare exact- and parent-ID gold-label coverage at candidate budgets of 10, 20, and 50. Concept-guided units and full candidate packets outperform raw-segment and evidence-span BM25 at each budget.}
    \caption{Candidate coverage at different retrieval budgets.}
    \label{fig:retrieval_coverage}
\end{figure}

\noindent\textbf{Evidence extraction.}
\emph{Raw segments + BM25}~\cite{robertson2009probabilistic} retrieves directly from report segments, whereas \emph{Evidence spans + BM25} uses the text of extracted observed evidence units as queries while keeping the retriever fixed. At a budget of 50, exact-ID coverage increases from 28.4\% to 31.6\%, and parent-level coverage from 39.3\% to 43.7\% (Figure~\ref{fig:retrieval_coverage}). At smaller budgets, the two BM25 variants achieve similar coverage. Evidence selection therefore provides a modest retrieval benefit, most evident at the larger candidate budget.

\noindent\textbf{Concept-based retrieval.}
For \emph{Concept-guided units}, we select primary candidates from the merged retrieval packets, keeping those linked to evidence units and supported by observed behavior. We rank this subset using their merged retrieval scores. At a budget of 50, this subset achieves 47.9\% exact-ID coverage and 60.7\% parent-level coverage. The \emph{Full candidate packet} additionally uses local evidence groups and ontology-based alternatives, increasing coverage to 52.5\% and 64.1\%. Both concept-based variants outperform the lexical retrievers at all three budgets and both granularities. These results indicate that concept-based retrieval yields a larger coverage gain than evidence selection alone. Additional retrieval-depth results are reported in Appendix~\ref{sec:retrieval_depth}.

\subsection{Output Control Ablation (RQ3)}
\label{sec:ablation}

\begingroup
\setlength{\emergencystretch}{1em}
\noindent\textbf{Effect of claim verification.}
Table~\ref{tab:decision_ablation} compares deduplicated claims accepted by the mapping LLM with the final decisions after document-level verification. Both settings use identical evidence, candidates, and mapping responses, isolating the effect of verification. The verifier yields 25 fewer exact-ID false positives and three fewer true positives, increasing macro precision from 0.366 to 0.384 and F1 from 0.306 to 0.310. Parent-level F1 also increases, from 0.383 to 0.386. Verification thus improves precision at the cost of a small reduction in recall, yielding a modest F1 gain.
\par
\endgroup

\begin{table}[t]
\centering
\caption{Effect of document-level output control.}
\label{tab:decision_ablation}
\begin{threeparttable}
    \renewcommand{\customsize}{\fontsize{7}{8.6}\selectfont}
    \setlength{\tabcolsep}{3.5pt}
    \renewcommand{\arraystretch}{1.00}
    \setlength{\aboverulesep}{1.5pt}
    \setlength{\belowrulesep}{1.5pt}
    \begin{smalltabularx}{.94\linewidth}{>{\centering\arraybackslash}p{9.5em} CCCC}
    \toprule[0.6pt]
    & \multicolumn{4}{c}{\textbf{Exact ATT\&CK IDs}} \\
    \cmidrule(lr){2-5}
    \textbf{Decision stage} & \textbf{F1} & \textbf{Recall} & \textbf{Prec.} & \textbf{Jacc.} \\
    \midrule[0.35pt]
    Mapping accepts only & 0.306 & \textbf{0.308} & 0.366 & 0.194 \\
    $+$ document verifier & \textbf{0.310} & 0.302 & \textbf{0.384} & \textbf{0.200} \\
    \midrule[0.3pt]
    & \multicolumn{4}{c}{\textbf{Parent IDs}} \\
    \cmidrule(lr){2-5}
    Mapping accepts only & 0.383 & \textbf{0.384} & 0.458 & 0.256 \\
    $+$ document verifier & \textbf{0.386} & 0.383 & \textbf{0.463} & \textbf{0.260} \\
    \bottomrule[0.6pt]
    \end{smalltabularx}

\end{threeparttable}
\end{table}



\begin{table*}[!t]
\centering
\caption{Technique identification performance with in-domain adaptation.}
\label{tab:trained_technique_identification}
\begin{threeparttable}
    \renewcommand{\customsize}{\fontsize{7}{8.6}\selectfont}
    \setlength{\tabcolsep}{3.5pt}
    \renewcommand{\arraystretch}{1.00}
    \setlength{\aboverulesep}{1.5pt}
    \setlength{\belowrulesep}{1.5pt}
    \begin{smalltabularx}{.94\linewidth}{c >{\centering\arraybackslash}p{9em} CCCC @{\hspace{8pt}} CCCC}
    \toprule[0.6pt]
    \multicolumn{2}{c}{\textbf{Method}} & \multicolumn{4}{c}{\textbf{Exact ATT\&CK IDs}} & \multicolumn{4}{c}{\textbf{Parent IDs}} \\
    \cmidrule(lr){1-2} \cmidrule(lr){3-6} \cmidrule(lr){7-10}
    \textbf{Type} & \textbf{Name} & \textbf{F1} & \textbf{Recall} & \textbf{Prec.} & \textbf{Jacc.} & \textbf{F1} & \textbf{Recall} & \textbf{Prec.} & \textbf{Jacc.} \\
    \midrule[0.35pt]
    \multirow{2}{*}{Classifier}
     & RoBERTa & 0.285 & 0.271 & \underline{0.328} & 0.183 & 0.358 & 0.365 & \underline{0.391} & 0.237 \\
     & CySecBERT & 0.199 & 0.187 & 0.242 & 0.119 & 0.306 & 0.313 & 0.343 & 0.195 \\
    \midrule[0.3pt]
    \multirow{3}{*}{SFT}
     & Llama-3.1-8B & 0.260 & 0.294 & 0.263 & 0.161 & 0.360 & 0.400 & 0.366 & 0.238 \\
     & Llama-3.1-70B & \underline{0.292} & \textbf{0.337} & 0.296 & \underline{0.187} & \underline{0.385} & \textbf{0.444} & 0.389 & \underline{0.260} \\
     & Qwen2.5-14B & 0.291 & \underline{0.331} & 0.293 & 0.184 & 0.368 & \underline{0.429} & 0.366 & 0.241 \\
    \midrule[0.3pt]
    \textbf{Ours} & \textbf{\sys{} + memory} & \textbf{0.316} & 0.317 & \textbf{0.379} & \textbf{0.205} & \textbf{0.394} & 0.396 & \textbf{0.465} & \textbf{0.269} \\
    \bottomrule[0.6pt]
    \end{smalltabularx}

\end{threeparttable}
\end{table*}

\noindent\textbf{Evidence behind decisions.}
In the Medusa case, the mapper accepts T1048 (Exfiltration Over Alternative Protocol) although the cited data-theft statement does not specify a transfer protocol. The verifier rejects the assignment because the candidate was retrieved only as a low-scoring alternative with no matched profile concepts. This removes one exact-ID false positive without changing the evidence or mapping response; the full comparison is shown in Figure~\ref{fig:verifier_case} in Appendix~\ref{sec:verifier_case_study}. 
We also report the corpus-wide structural audit in Appendix~\ref{sec:traceability_audit} and discuss potential analyst workflows using these records in Appendix~\ref{sec:analyst_workflows}.

\subsection{Evidence-Memory Adaptation (RQ4)}
\label{sec:evidence_memory_results}

We evaluate optional evidence-memory adaptation on the same 55 reports used in RQ1. Table~\ref{tab:trained_technique_identification} compares this variant against classifiers and LLM baselines adapted using the training partition. Without updating model parameters, \sys{} with memory achieves exact-ID and parent-level macro F1 scores of 0.316 and 0.394, compared with 0.292 and 0.385 for the strongest adapted baseline. 



We apply the evidence-memory revision described in Section~\ref{sec:evidence_memory}. For paired evaluation, we rerun mapping for revised packets, retain new claims only for promoted and demoted techniques, and reuse archived claims for all other techniques. Memory revision modifies 66 packets without increasing their decision budgets and increases the number of gold report--technique associations included in the decision queues from 264 to 271. Final predictions show a net gain of seven exact-ID true positives, with macro F1 rising from 0.310 to 0.316. Parent-level macro F1 likewise increases from 0.386 to 0.394.

\begin{figure}[t]
    \centering
    \includegraphics[width=\linewidth]{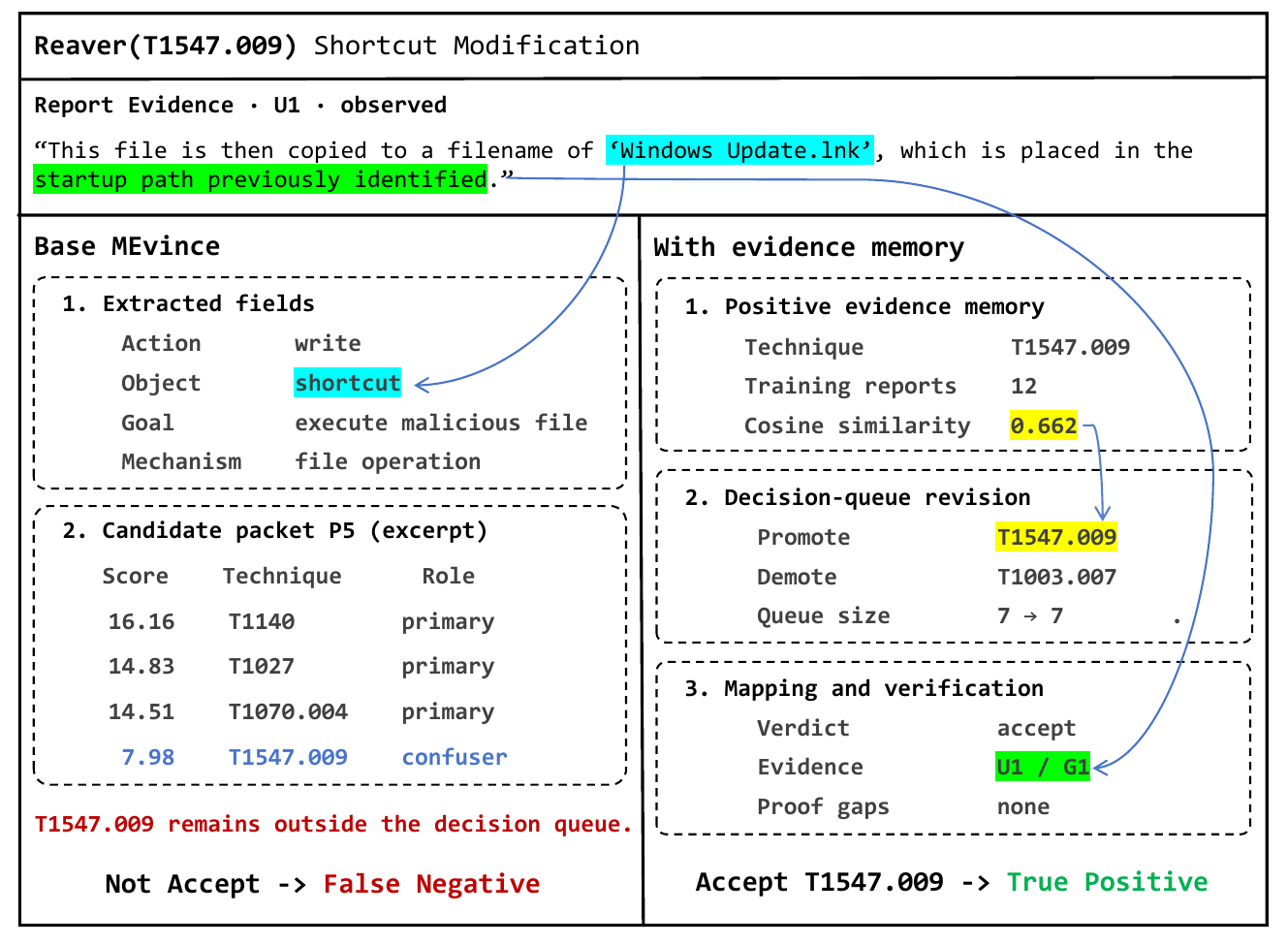}
    \Description{A paired Reaver case study shows that evidence memory promotes T1547.009 Shortcut Modification from a confuser outside the mandatory decision queue into the same-size queue. The resulting acceptance cites evidence U1 and group G1 from the current report.}
    \caption{Evidence-memory revision for T1547.009.}
    \label{fig:training_case}
    \vspace{-5pt}
\end{figure}

Figure~\ref{fig:training_case} illustrates this mechanism in the Reaver report. The report describes malware creating a shortcut and placing it in a previously identified startup path. \sys{} initially retrieves T1547.009 as a confuser but leaves it outside the mandatory decision queue. Evidence prototypes from 12 training reports promote it into the queue by replacing one incumbent, keeping the queue at seven candidates. The resulting acceptance cites U1 and G1 from the current report and has no proof gap. At the report level, adaptation recovers T1547.009 and T1012 without adding a false positive. 

\subsection{LLM Workload (RQ5)}
\label{sec:llm_workload}

We compare the online LLM workload of Direct Prompt, RAG, CTIKG, and \sys{} on the shortest, median-length, and longest reports among the 55 evaluation reports, selected by character length. All four methods use GPT-4o. Table~\ref{tab:llm_cost} reports generation calls and total token usage recorded by the API, excluding embedding queries and offline preparation. 
Direct Prompt uses one generation call per report and has the lowest token consumption. \sys{} uses more tokens than RAG, as it runs separate LLM stages for evidence extraction and technique mapping. Compared with CTIKG, it requires fewer calls and reduces generation-token usage by 62.1--83.8\%.
Input/output breakdowns are provided in Table~\ref{tab:llm_token_breakdown}.
Across the evaluation set, estimated generation-token usage for \sys{} averages 49.9K per report, with a median of 41.8K.

\begin{table}[t]
\centering
\vspace{-3pt}
\caption{Online LLM workload across report lengths.}
\label{tab:llm_cost}
\begin{threeparttable}
\footnotesize
\setlength{\tabcolsep}{0pt}
\renewcommand{\arraystretch}{1.15}
\begin{tabular*}{\linewidth}{@{\extracolsep{\fill}}l@{\hspace{7pt}}r@{\hspace{6pt}}r@{\hspace{12pt}}r@{\hspace{6pt}}r@{\hspace{12pt}}r@{\hspace{6pt}}r@{}}
\toprule
& \multicolumn{2}{c}{\textbf{Shortest}} & \multicolumn{2}{c}{\textbf{Median}} & \multicolumn{2}{c}{\textbf{Longest}} \\
\cmidrule(lr){2-3}\cmidrule(lr){4-5}\cmidrule(lr){6-7}
\textbf{Method} & \textbf{Calls} & \textbf{Tokens (K)} & \textbf{Calls} & \textbf{Tokens (K)} & \textbf{Calls} & \textbf{Tokens (K)} \\
\midrule
Direct Prompt & 1 & 0.97 & 1 & 3.55 & 1 & 10.11 \\
RAG & 4 & 4.25 & 24 & 25.26 & 72 & 79.01 \\
CTIKG & 39 & 35.01 & 249 & 254.30 & 794 & 791.34 \\
\midrule[0.3pt]
\textbf{\sys{}} & 6 & 13.28 & 12 & 41.29 & 45 & 144.11 \\
\bottomrule
\end{tabular*}
\begin{tablenotes}
\scriptsize
\item[] Report lengths of the three settings are 2,059, 11,238, and 34,870 characters, respectively.
\end{tablenotes}
\end{threeparttable}
\end{table}

\section{Conclusion}
We presented \sys{}, a knowledge-constrained LLM reasoning framework for report-level malware technique identification.
The framework aligns dispersed report observations with ATT\&CK technique definitions through a shared behavioral representation, while maintaining a clear distinction among source evidence, interpretive knowledge, and unverified hypotheses.
Evaluations against 11 baselines on public malware reports demonstrate that \sys{} improves fine-grained technique identification, while an offline evidence memory enables further gains by refining bounded candidate queues without updating model parameters.
Further analyses show that the framework improves candidate coverage while reducing false-positive assignments, and provenance audits confirm that accepted assignments remain traceable to source evidence.


\bibliographystyle{ACM-Reference-Format}
\bibliography{reference}

@misc{mitre_attack,
  author={{The MITRE Corporation}},
  title={ATT\&CK Matrix for Enterprise},
  howpublished = {\url{https://attack.mitre.org}},
  note={Accessed: 2026-03-29}
}

@inproceedings{husari2017ttpdrill,
  title={TTPDrill: Automatic and accurate extraction of threat actions from unstructured text of cti sources},
  author={Husari, Ghaith and Al-Shaer, Ehab and Ahmed, Mohiuddin and Chu, Bill and Niu, Xi},
  booktitle={Proceedings of the 33rd annual computer security applications conference},
  pages={103--115},
  year={2017}
}

@inproceedings{guru2025technique,
  title={On Technique Identification and Threat-Actor Attribution using LLMs and Embedding Models},
  author={Guru, Kyla and Moss, Robert J and Kochenderfer, Mykel J},
  booktitle={2025 International Conference on Cybersecurity and AI-Based Systems (Cyber-AI)},
  pages={332--339},
  year={2025},
  organization={IEEE}
}

@article{legoy2020rcatt,
  title={Automated retrieval of att\&ck tactics and techniques for cyber threat reports},
  author={Legoy, Valentine and Caselli, Marco and Seifert, Christin and Peter, Andreas},
  journal={arXiv preprint arXiv:2004.14322},
  year={2020}
}

@inproceedings{buchel2025sok,
title={{SoK}: Automated {TTP} Extraction from {CTI} Reports--Are We There Yet?},
  author={B{\"u}chel, Marvin and Paladini, Tommaso and Longari, Stefano and Carminati, Michele and Zanero, Stefano and Binyamini, Hodaya and Engelberg, Gal and Klein, Dan and Guizzardi, Giancarlo and Caselli, Marco and others},
  booktitle={34th USENIX security symposium (USENIX Security 25)},
  pages={4621--4641},
  year={2025}
}

@inproceedings{li2022attackg,
  title={AttacKG: Constructing technique knowledge graph from cyber threat intelligence reports},
  author={Li, Zhenyuan and Zeng, Jun and Chen, Yan and Liang, Zhenkai},
  booktitle={European symposium on research in computer security},
  pages={589--609},
  year={2022},
  organization={Springer}
}

@article{liu2022threat,
  title={Threat intelligence ATT\&CK extraction based on the attention transformer hierarchical recurrent neural network},
  author={Liu, Chenjing and Wang, Junfeng and Chen, Xiangru},
  journal={Applied Soft Computing},
  volume={122},
  pages={108826},
  year={2022},
  publisher={Elsevier}
}

@article{rani2024ttpxhunter,
  title={TTPXHunter: Actionable threat intelligence extraction as ttps from finished cyber threat reports},
  author={Rani, Nanda and Saha, Bikash and Maurya, Vikas and Shukla, Sandeep Kumar},
  journal={Digital Threats: Research and Practice},
  volume={5},
  number={4},
  pages={1--19},
  year={2024},
  publisher={ACM New York, NY}
}

@inproceedings{rahman2024alert,
  title={Alert: A framework for efficient extraction of attack techniques from cyber threat intelligence reports using active learning},
  author={Rahman, Fariha Ishrat and Halim, Sadaf Md and Singhal, Anoop and Khan, Latifur},
  booktitle={IFIP Annual Conference on Data and Applications Security and Privacy},
  pages={203--220},
  organization={Springer},
  year={2024},
}

@article{cheng2024ctinexus,
  title={CTINEXUS: leveraging optimized LLM in-context learning for constructing cybersecurity knowledge graphs under data scarcity},
  author={Cheng, Yutong and Bajaber, Osama and Tsegai, Saimon Amanuel and Song, Dawn and Gao, Peng},
  journal={arXiv preprint arXiv:2410.21060},
  year={2024}
}

@inproceedings{ranade2021cybert,
  title={Cybert: Contextualized embeddings for the cybersecurity domain},
  author={Ranade, Priyanka and Piplai, Aritran and Joshi, Anupam and Finin, Tim},
  booktitle={2021 IEEE international conference on big data (Big Data)},
  pages={3334--3342},
  year={2021},
  organization={IEEE}
}

@article{bayer2024cysecbert,
  title={Cysecbert: A domain-adapted language model for the cybersecurity domain},
  author={Bayer, Markus and Kuehn, Philipp and Shanehsaz, Ramin and Reuter, Christian},
  journal={ACM Transactions on Privacy and Security},
  volume={27},
  number={2},
  pages={1--20},
  year={2024},
  publisher={ACM New York, NY}
}

@inproceedings{jin2023darkbert,
  title={Darkbert: A language model for the dark side of the internet},
  author={Jin, Youngjin and Jang, Eugene and Cui, Jian and Chung, Jin-Woo and Lee, Yongjae and Shin, Seungwon},
  booktitle={Proceedings of the 61st annual meeting of the association for computational linguistics (volume 1: long papers)},
  pages={7515--7533},
  year={2023}
}

@article{chen2025aecr,
  title={AECR: Automatic attack technique intelligence extraction based on fine-tuned large language model},
  author={Chen, Minghao and Zhu, Kaijie and Lu, Bin and Li, Ding and Yuan, Qingjun and Zhu, Yuefei},
  journal={Computers \& Security},
  volume={150},
  pages={104213},
  year={2025},
  publisher={Elsevier}
}

@inproceedings{huang2024ctikg,
  title={CTIKG: LLM-Powered Knowledge Graph Construction from Cyber Threat Intelligence},
  author={Liangyi Huang and Xusheng Xiao},
  booktitle={Proceedings of the First Conference on Language Modeling (COLM 2024)},
  year={2024},
  location={Philadelphia, PA, USA},
  month={October}
}

@inproceedings{fayyazi2024advancing,
  title={Advancing TTP analysis: Harnessing the power of large language models with retrieval augmented generation},
  author={Fayyazi, Reza and Taghdimi, Rozhina and Yang, Shanchieh Jay},
  booktitle={2024 Annual Computer Security Applications Conference Workshops (ACSAC Workshops)},
  pages={255--261},
  year={2024},
  organization={IEEE}
}

@inproceedings{fieblinger2024actionable,
  title={Actionable cyber threat intelligence using knowledge graphs and large language models},
  author={Fieblinger, Romy and Alam, Md Tanvirul and Rastogi, Nidhi},
  booktitle={2024 IEEE European symposium on security and privacy workshops (EuroS\&PW)},
  pages={100--111},
  year={2024},
  organization={IEEE}
}

@inproceedings{guo2025intelliradar,
  title={IntelliRadar: A Comprehensive Platform to Pinpoint Malicious Package Information from Cyber Intelligence},
  author={Wenbo Guo and Chengwei Liu and Limin Wang and Yiran Zhang and Jiahui Wu and Zhengzi Xu and Yang Liu},
  booktitle={2026 IEEE/ACM 48th International Conference on Software Engineering (ICSE)},
  year={2026},
  organization={IEEE Computer Society}
}

@inproceedings{liao2016acing,
  title={Acing the ioc game: Toward automatic discovery and analysis of open-source cyber threat intelligence},
  author={Liao, Xiaojing and Yuan, Kan and Wang, XiaoFeng and Li, Zhou and Xing, Luyi and Beyah, Raheem},
  booktitle={Proceedings of the 2016 ACM SIGSAC conference on computer and communications security},
  pages={755--766},
  year={2016}
}

@article{grattafiori2024llama,
  title={The llama 3 herd of models},
  author={Grattafiori, Aaron and Dubey, Abhimanyu and Jauhri, Abhinav and Pandey, Abhinav and Kadian, Abhishek and Al-Dahle, Ahmad and Letman, Aiesha and Mathur, Akhil and Schelten, Alan and Vaughan, Alex and others},
  journal={arXiv preprint arXiv:2407.21783},
  year={2024}
}

@article{yang2024qwen2,
  title={Qwen2 Technical Report},
  author={Yang, An and Yang, Baosong and Hui, Binyuan and Zheng, Bo and Yu, Bowen and Zhou, Chang and Li, Chengpeng and Li, Chengyuan and Liu, Dayiheng and Huang, Fei and others},
  journal={arXiv preprint arXiv:2407.10671},
  year={2024},
  url={https://arxiv.org/abs/2407.10671}
}

@article{li2022nersurvey,
  title={A survey on deep learning for named entity recognition},
  author={Li, Jing and Sun, Aixin and Han, Jianglei and Li, Chenliang},
  journal={IEEE transactions on knowledge and data engineering},
  volume={34},
  number={1},
  pages={50--70},
  year={2022},
  publisher={IEEE}
}

@misc{ross2023tram2,
  author={James Ross and Jackie Lasky},
  title={Our TRAM Large Language Model Automates TTP Identification in CTI Reports},
  month=Aug,
  howpublished={\url{https://medium.com/mitre-engenuity/our-tram-large-language-model-automates-ttp-identification-in-cti-reports-5bc0a30d4567}},
  note={MITRE Engenuity Center for Threat-Informed Defense, published August 29, 2023; accessed March 31, 2026},
  year={2023},
}

@inproceedings{he2025combating,
  title={Combating concept drift with explanatory detection and adaptation for android malware classification},
  author={He, Yiling and Lei, Junchi and Qin, Zhan and Ren, Kui and Chen, Chun},
  booktitle={Proceedings of the 2025 ACM SIGSAC Conference on Computer and Communications Security},
  pages={978--992},
  year={2025}
}

@inproceedings{chen2023learning,
  title={Learning implicit entity-object relations by bidirectional generative alignment for multimodal ner},
  author={Chen, Feng and Liu, Jiajia and Ji, Kaixiang and Ren, Wang and Wang, Jian and Chen, Jingdong},
  booktitle={Proceedings of the 31st ACM international conference on multimedia},
  pages={4555--4563},
  year={2023}
}

@inproceedings{perera2015implicit,
  title={Implicit entity recognition in clinical documents},
  author={Perera, Sujan and Mendes, Pablo and Sheth, Amit and Thirunarayan, Krishnaprasad and Alex, Adarsh and Heid, Christopher and Mott, Greg},
  booktitle={Proceedings of the Fourth Joint Conference on Lexical and Computational Semantics},
  pages={228--238},
  year={2015}
}

@misc{trendmicro2018njrat,
  author={Pascual, Carl Maverick and Ofiaza, Michael Jhon and Roderno, Patrick Angelo},
  title ={AutoIt-Compiled Worm Sends Fileless BLADABINDI/njRAT},
  year={2018},
  month=nov,
  howpublished={\url{https://www.trendmicro.com/en_us/research/18/k/autoit-compiled-worm-affecting-removable-media-delivers-fileless-version-of-bladabindi-njrat-backdoor.html}},
  note={Trend Micro Research, accessed March 31, 2026}
}

@inproceedings{sharma2023ttp,
  title={To ttp or not to ttp?: Exploiting ttps to improve ml-based malware detection},
  author={Sharma, Yashovardhan and Giunchiglia, Eleonora and Birnbach, Simon and Martinovic, Ivan},
  booktitle={2023 IEEE International Conference on Cyber Security and Resilience (CSR)},
  pages={8--15},
  year={2023},
  organization={IEEE}
}

@article{xuan2026identifying,
  title={Identifying Adversary Tactics and Techniques in Malware Binaries with an LLM Agent},
  author={Xuan, Zhou and Xu, Xiangzhe and Zheng, Mingwei and Tan, Louis Zheng-Hua and Guo, Jinyao and Zhang, Tiantai and Yu, Le and Wang, Chengpeng and Zhang, Xiangyu},
  journal={arXiv preprint arXiv:2602.06325},
  year={2026}
}

@inproceedings{ayoade2018automated,
  title={Automated threat report classification over multi-source data},
  author={Ayoade, Gbadebo and Chandra, Swarup and Khan, Latifur and Hamlen, Kevin and Thuraisingham, Bhavani},
  booktitle={2018 IEEE 4th International Conference on Collaboration and Internet Computing (CIC)},
  pages={236--245},
  year={2018},
  organization={IEEE}
}

@article{satvat2021extractor,
  title={Extractor: Extracting attack behavior from threat reports},
  author={Satvat, Kiavash and Gjomemo, Rigel and Venkatakrishnan, VN},
  journal={arXiv preprint arXiv:2104.08618},
  year={2021}
}

@inproceedings{husari2018actionminer,
  title={Using entropy and mutual information to extract threat actions from cyber threat intelligence},
  author={Husari, Ghaith and Niu, Xi and Chu, Bill and Al-Shaer, Ehab},
  booktitle={2018 IEEE international conference on intelligence and security informatics (ISI)},
  pages={1--6},
  year={2018},
  organization={IEEE}
}

@article{zhang2021ex,
  title={EX-Action: Automatically Extracting Threat Actions from Cyber Threat Intelligence Report Based on Multimodal Learning},
  author={Zhang, Huixia and Shen, Guowei and Guo, Chun and Cui, Yunhe and Jiang, Chaohui},
  journal={Security and Communication Networks},
  volume={2021},
  number={1},
  pages={5586335},
  year={2021},
  publisher={Wiley Online Library}
}

@article{furumoto2025survey,
  title={A Comprehensive Survey of Threat Intelligence Research: A Measurement-Based Study},
  author={Furumoto, Keisuke and Morikawa, Tomohiro and Kolehmainen, Antti and Silverajan, Bilhanan and Takahashi, Takeshi and Inoue, Daisuke},
  journal={ACM Computing Surveys},
  volume={58},
  number={6},
  articleno={153},
  numpages={35},
  year={2025},
  month=dec,
  publisher={Association for Computing Machinery},
  doi={10.1145/3772280}
}

@article{qwen2025qwen25,
  title={{Qwen2.5} Technical Report},
  author={Qwen and Yang, An and Yang, Baosong and Zhang, Beichen and Hui, Binyuan and Zheng, Bo and Yu, Bowen and Li, Chengyuan and Liu, Dayiheng and Huang, Fei and others},
  journal={arXiv preprint arXiv:2412.15115},
  year={2025},
  url={https://arxiv.org/abs/2412.15115}
}

@inproceedings{lewis2020retrieval,
  title={Retrieval-Augmented Generation for Knowledge-Intensive {NLP} Tasks},
  author={Lewis, Patrick and Perez, Ethan and Piktus, Aleksandra and Petroni, Fabio and Karpukhin, Vladimir and Goyal, Naman and K{\"u}ttler, Heinrich and Lewis, Mike and Yih, Wen-tau and Rockt{\"a}schel, Tim and Riedel, Sebastian and Kiela, Douwe},
  booktitle={Advances in Neural Information Processing Systems},
  volume={33},
  pages={9459--9474},
  publisher={Curran Associates, Inc.},
  year={2020},
  url={https://proceedings.neurips.cc/paper/2020/hash/6b493230205f780e1bc26945df7481e5-Abstract.html}
}

@article{liu2019roberta,
  title={{RoBERTa}: A Robustly Optimized {BERT} Pretraining Approach},
  author={Liu, Yinhan and Ott, Myle and Goyal, Naman and Du, Jingfei and Joshi, Mandar and Chen, Danqi and Levy, Omer and Lewis, Mike and Zettlemoyer, Luke and Stoyanov, Veselin},
  journal={arXiv preprint arXiv:1907.11692},
  year={2019},
  url={https://arxiv.org/abs/1907.11692}
}

@article{robertson2009probabilistic,
  title={The Probabilistic Relevance Framework: {BM25} and Beyond},
  author={Robertson, Stephen and Zaragoza, Hugo},
  journal={Foundations and Trends in Information Retrieval},
  volume={3},
  number={4},
  pages={333--389},
  year={2009},
  doi={10.1561/1500000019}
}

@article{douze2024faiss,
  title={The {Faiss} library},
  author={Douze, Matthijs and Guzhva, Alexandr and Deng, Chengqi and Johnson, Jeff and Szilvasy, Gergely and Mazar{\'e}, Pierre-Emmanuel and Lomeli, Maria and Hosseini, Lucas and J{\'e}gou, Herv{\'e}},
  journal={arXiv preprint arXiv:2401.08281},
  year={2024},
  url={https://arxiv.org/abs/2401.08281}
}

@article{kalajdzievski2023rank,
  title={A Rank Stabilization Scaling Factor for Fine-Tuning with {LoRA}},
  author={Kalajdzievski, Damjan},
  journal={arXiv preprint arXiv:2312.03732},
  year={2023},
  url={https://arxiv.org/abs/2312.03732}
}

@inproceedings{devlin2019bert,
  title={{BERT}: Pre-training of Deep Bidirectional Transformers for Language Understanding},
  author={Devlin, Jacob and Chang, Ming-Wei and Lee, Kenton and Toutanova, Kristina},
  booktitle={Proceedings of the 2019 Conference of the North American Chapter of the Association for Computational Linguistics: Human Language Technologies, Volume 1 (Long and Short Papers)},
  pages={4171--4186},
  year={2019},
  publisher={Association for Computational Linguistics},
  doi={10.18653/v1/N19-1423},
  url={https://aclanthology.org/N19-1423/}
}

@article{efron1979bootstrap,
  title={Bootstrap Methods: Another Look at the Jackknife},
  author={Efron, B.},
  journal={The Annals of Statistics},
  volume={7},
  number={1},
  pages={1--26},
  year={1979},
  doi={10.1214/aos/1176344552}
}

@inproceedings{gao2021enabling,
  title={Enabling Efficient Cyber Threat Hunting With Cyber Threat Intelligence},
  author={Gao, Peng and Shao, Fei and Liu, Xiaoyuan and Xiao, Xusheng and Qin, Zheng and Xu, Fengyuan and Mittal, Prateek and Kulkarni, Sanjeev R. and Song, Dawn},
  booktitle={2021 IEEE 37th International Conference on Data Engineering (ICDE)},
  pages={193--204},
  year={2021},
  publisher={IEEE},
  doi={10.1109/ICDE51399.2021.00024}
}

@inproceedings{gao2023citations,
  title={Enabling Large Language Models to Generate Text with Citations},
  author={Gao, Tianyu and Yen, Howard and Yu, Jiatong and Chen, Danqi},
  booktitle={Proceedings of the 2023 Conference on Empirical Methods in Natural Language Processing},
  pages={6465--6488},
  year={2023},
  publisher={Association for Computational Linguistics},
  doi={10.18653/v1/2023.emnlp-main.398},
  url={https://aclanthology.org/2023.emnlp-main.398/}
}

@inproceedings{gao2023rarr,
  title={{RARR}: Researching and Revising What Language Models Say, Using Language Models},
  author={Gao, Luyu and Dai, Zhuyun and Pasupat, Panupong and Chen, Anthony and Chaganty, Arun Tejasvi and Fan, Yicheng and Zhao, Vincent and Lao, Ni and Lee, Hongrae and Juan, Da-Cheng and Guu, Kelvin},
  booktitle={Proceedings of the 61st Annual Meeting of the Association for Computational Linguistics (Volume 1: Long Papers)},
  pages={16477--16508},
  year={2023},
  publisher={Association for Computational Linguistics},
  doi={10.18653/v1/2023.acl-long.910},
  url={https://aclanthology.org/2023.acl-long.910/}
}

\appendix

\section{Supplementary Discussion}
\label{sec:supplementary_discussion}
\noindent\textbf{Evidence Control and the Precision--Recall Tradeoff.}
RAG attains high recall but produces many assignments that do not match the benchmark associations; \sys{} instead exposes uncertain claims as abstentions. The retrieval ablation shows that most of the candidate-coverage improvement comes from comparing evidence and profiles in the shared concept space, with a smaller additional gain from evidence extraction. At the decision stage, document-level verification provides a modest precision gain on identical mapping responses.
The evidence-memory experiment shows that positive training examples can improve candidate selection before mapping. The memory improves recall on the 55 evaluation reports while retaining the same mapping and verification procedures.
In multi-label mapping, every weak assignment added for recall becomes another claim to verify before it can safely enter a knowledge base or detection workflow. Retrieval should preserve plausible coverage, while acceptance remains conservative about what the report supports; abstention keeps a candidate for review without presenting it as observed behavior.

\noindent\textbf{Traceability and Analyst Interaction.}
Accepted and abstained claims link through evidence units and groups to source spans. Candidate packets retain matched concepts and confusable alternatives, while claim records retain evidence references and proof gaps. These records help an analyst distinguish extraction errors, retrieval mismatches, and unsupported decisions. The claim score summarizes how well the evidence satisfies the verification rules. It is not a calibrated probability; consequential use still requires human review.

\noindent\textbf{Limitations.}
The development split informed method design. A larger independent evaluation, especially on unseen software families, is needed. Moreover, ATT\&CK labels are scoped to cited software--report associations rather than exhaustive behavior annotations. Label disagreement therefore does not by itself establish an unsupported claim. Our provenance checks establish that citations resolve to report spans, not that an expert would judge each span sufficient for its technique; evaluating that judgment and analyst utility remains future work.

\section{Concept Inventory and Semantic Composition}
\label{sec:concept_examples}

Table~\ref{tab:concept_examples} summarizes the inventory used in our experiments, with representative entries for each component. The four concept dimensions contain 450 canonical behavioral labels. Each label has a definition and lexical or structural anchors; these primitives describe behavior rather than assign technique identifiers. The semantic layer contains 34 role templates and 17 coupling motifs. A role template combines primitives to describe a behavior, such as registry persistence. A coupling motif links behaviors described in nearby observations, such as delivery followed by execution.

\begin{table*}[t]
\caption{Concept inventory used by \sys{}.}
\label{tab:concept_examples}
\centering
\footnotesize
\setlength{\tabcolsep}{6pt}
\renewcommand{\arraystretch}{1.15}
\renewcommand{\tabularxcolumn}[1]{m{#1}}
\begin{tabularx}{.94\linewidth}{l r >{\raggedright\arraybackslash}X >{\raggedright\arraybackslash}X}
\toprule
\textbf{Component} & \textbf{Count} & \textbf{Role} & \textbf{Representative entries} \\
\midrule
\textbf{Action} & 36 & What the threat does; preserves the behavioral predicate. & execute; modify; query; inject; connect \\
\textbf{Object} & 94 & What the action affects; identifies the artifact or resource. & process; file; registry key; credential; network configuration \\
\textbf{Goal} & 18 & Why the behavior is performed; separates similar operations with different purposes. & persistence; discovery; credential access; command and control \\
\textbf{Mechanism} & 302 & How the behavior is carried out; captures technique-specific means. & Run-key startup; PowerShell; process hollowing; non-standard port \\
\midrule
\textbf{Role template} & 34 & Combines concepts that describe the same behavior. & registry persistence; fileless execution; discovery; C2; exfiltration \\
\textbf{Coupling motif} & 17 & Connects related behaviors described in nearby evidence units. & delivery--execution; Run-key--fileless execution; C2--exfiltration \\
\bottomrule
\end{tabularx}
\end{table*}

\section{Reproducibility Details}
\label{sec:reproducibility}

\begingroup
\setlength{\emergencystretch}{2em}

\sys{} uses the ATT\&CK Enterprise v19.2 bundle and the \texttt{gpt-4o} alias with temperature 1.0 and a 2,048-token output limit. Base profiles use ATT\&CK technique IDs, names, and descriptions, not software--technique usage records. Baselines retain their method-specific prompts and decoding schedules. All reported metrics are computed from per-report predictions using the common evaluator.

\noindent\textbf{Evidence-memory configuration.} The memory contains 9,440 positive prototypes from the 555-report training partition. For each technique annotated in a training report, we select evidence units that describe observed behavior and are relevant to mapping. We rank these units by their TF--IDF alignment with the technique profile and retain at most two with scores of at least 0.01. Evidence texts and prototypes are encoded with \texttt{text-embedding-3-small} at 512 dimensions. Configuration selection uses five-fold validation grouped by software identifier, with prototypes from each held-out fold excluded. On held-out training reports, we measure how many gold candidates each pre-specified configuration adds to or removes from the base queue. We select the configuration with the largest net gain per revised packet. The selected setting uses a cosine threshold of 0.65, requires prototypes from at least five training reports per eligible technique, and permits at most two exchanges per packet. The memory and configuration are fixed before evaluation.

\noindent\textbf{Archived outputs and paired evaluation.} Each report retains separate JSON outputs for extraction, concept mining, retrieval, mapping, and final decisions. These records preserve evidence units and groups, candidate packets, mapping responses, and decision references. For evidence-memory evaluation, we rerun mapping for revised packets, retain new claims only for promoted and demoted techniques, and reuse archived claims for all other techniques. Both variants use the same extraction results and document-level verifier.

\noindent\textbf{Component comparisons.} Retrieval coverage is evaluated on the same report inputs and technique profiles, using the same report-level candidate budgets for all variants. The output-control comparison holds evidence, candidates, and mapping responses fixed, changing only document-level verification.

\noindent\textbf{Workload accounting.} The comparison in Section~\ref{sec:llm_workload} uses API-recorded input and output tokens from separate replays on reports 2, 323, and 705, the shortest, median-length, and longest reports by character count in the 55-report evaluation set. Table~\ref{tab:llm_token_breakdown} provides the input/output breakdown. Generation calls are counted separately from embedding queries, and offline preparation is excluded. RAG uses 4, 24, and 72 embedding queries on these reports, corresponding to 404, 2,454, and 8,751 input-token equivalents, respectively. For the 55-report estimates, extraction and mapping prompts are reconstructed from recorded inputs using the original prompt templates, and parsed responses are serialized consistently before tokenization. Replay outputs are not used to compute identification accuracy.

\begin{table}[t]
\centering
\caption{Input and output token usage.}
\label{tab:llm_token_breakdown}
\begin{threeparttable}
\footnotesize
\setlength{\tabcolsep}{3pt}
\renewcommand{\arraystretch}{1.08}
\begin{tabular*}{.94\linewidth}{@{\extracolsep{\fill}}lrrrrrr@{}}
\toprule
& \multicolumn{2}{c}{\textbf{Shortest}} & \multicolumn{2}{c}{\textbf{Median}} & \multicolumn{2}{c}{\textbf{Longest}} \\
\cmidrule(lr){2-3}\cmidrule(lr){4-5}\cmidrule(lr){6-7}
\textbf{Method} & \textbf{Input} & \textbf{Output} & \textbf{Input} & \textbf{Output} & \textbf{Input} & \textbf{Output} \\
\midrule
Direct Prompt & 0.65 & 0.32 & 2.91 & 0.64 & 9.18 & 0.93 \\
RAG & 3.81 & 0.44 & 22.92 & 2.34 & 70.90 & 8.11 \\
CTIKG & 28.54 & 6.47 & 206.02 & 48.28 & 648.18 & 143.16 \\
\midrule[0.3pt]
\textbf{\sys{}} & 9.62 & 3.66 & 28.49 & 12.81 & 99.25 & 44.86 \\
\bottomrule
\end{tabular*}
\begin{tablenotes}
\scriptsize
\item[] Values are in thousands of generation tokens, rounded independently. The reports are the same as those in Table~\ref{tab:llm_cost}.
\end{tablenotes}
\end{threeparttable}
\end{table}

\par
\endgroup

\subsection{Baseline Descriptions and Adaptation}
\label{sec:baseline_details}

\noindent\textbf{Classification.} RoBERTa and CySecBERT provide encoder-based classification baselines. RoBERTa uses general-purpose language representations, whereas CySecBERT incorporates continued pretraining on cybersecurity text~\cite{bayer2024cysecbert}. Their report-level technique predictions are evaluated using the common evaluator.

\noindent\textbf{NLP-based extraction and mapping.} We follow the NER pipeline designs studied by B{\"u}chel et al.~\cite{buchel2025sok}. NER-base performs exact matching; NER-full additionally uses part-of-speech tagging, lemmatization, related-word detection, and parsing to recognize explicit references despite lexical and syntactic variation. These pipelines do not require labeled task-specific training data. AttacKG~\cite{li2022attackg} extracts attack-relevant entities and dependencies, constructs behavior graphs, and identifies techniques by aligning these graphs with technique templates. We retain the technique identifiers returned by each pipeline, including parent-level predictions.

\noindent\textbf{Supervised fine-tuning.} The SFT baselines use 8B and 70B Llama instruction models and a 14B Qwen instruction model. Sentence-level training converts labeled text into chat-formatted examples asking the model to identify techniques expressed in each sentence. We use Unsloth with rank-stabilized LoRA~\cite{kalajdzievski2023rank}, rank 64, an alpha of 16, and zero adapter dropout. Adapters target attention and feed-forward projections, token embeddings, and the output head. The shared recipe uses three epochs, a per-device batch size of two, and two gradient-accumulation steps; the backbone and learning rate vary by configuration. At inference, each sentence is processed with greedy decoding and at most 256 new tokens. Technique identifiers are extracted from the responses and deduplicated across sentences to obtain report-level predictions.

\noindent\textbf{Context-driven LLM methods.} Direct Prompt, RAG, and CTIKG use GPT-4o for inference. Direct Prompt supplies the complete report and requests technique identifiers with supporting report quotes, without external retrieval. Our RAG implementation combines BM25 and FAISS-based dense retrieval~\cite{douze2024faiss} with weights of 0.3 and 0.7, respectively, using \texttt{text-embedding-3-large}. It indexes non-deprecated, Windows-relevant Enterprise technique documentation. For each report segment, the LLM receives up to ten deduplicated technique descriptions, each truncated to 240 characters, and identifies techniques supported by that segment. Predictions are combined across segments. Our CTIKG implementation~\cite{huang2024ctikg} performs repeated triple extraction, consolidates entities and relations into report-level memory, and maps the resulting representation to technique identifiers.

\noindent\textbf{Adaptation and evaluation.} Table~\ref{tab:technique_identification} uses baseline configurations without additional fitting on our 555-report training partition; some underlying models have been trained on other data. Table~\ref{tab:trained_technique_identification} separately evaluates classifiers and SFT models adapted on that partition, alongside \sys{} with evidence memory. Direct Prompt, RAG, and CTIKG receive no additional fitting. Both comparisons use the same 55 evaluation reports. Methods retain their own text processing and prediction rules; the common evaluator parses and deduplicates report-level technique identifiers and applies the same metrics.

\subsection{Retrieval Depth}
\label{sec:retrieval_depth}

Without the report-level budget used in Figure~\ref{fig:retrieval_coverage}, the full candidate packets cover 329 of 503 exact-ID report--technique associations (65.4\%) and 371 of 471 parent-level associations (78.8\%). Increasing the per-target retrieval depth from four to eight raises coverage to 66.0\% and 80.3\%, respectively, while the mean number of distinct candidates per report grows from 109.5 to 115.5. Increasing retrieval depth alone therefore recovers few additional reference labels.

\subsection{Structural Traceability Audit}
\label{sec:traceability_audit}

We audit all 438 report--technique assignments accepted by the frozen base run on the 55 evaluation reports. Each has a supporting claim, at least one resolvable evidence unit or group, and rendered evidence text. None of their supporting claims records a blocking proof gap. This audit verifies that the required evidence references and decision records are present and resolvable; it does not evaluate whether the cited behavior is sufficient to establish the technique.

\subsection{Document-Verification Case Study}
\label{sec:verifier_case_study}

\begin{figure}[t]
    \centering
    \includegraphics[width=\linewidth]{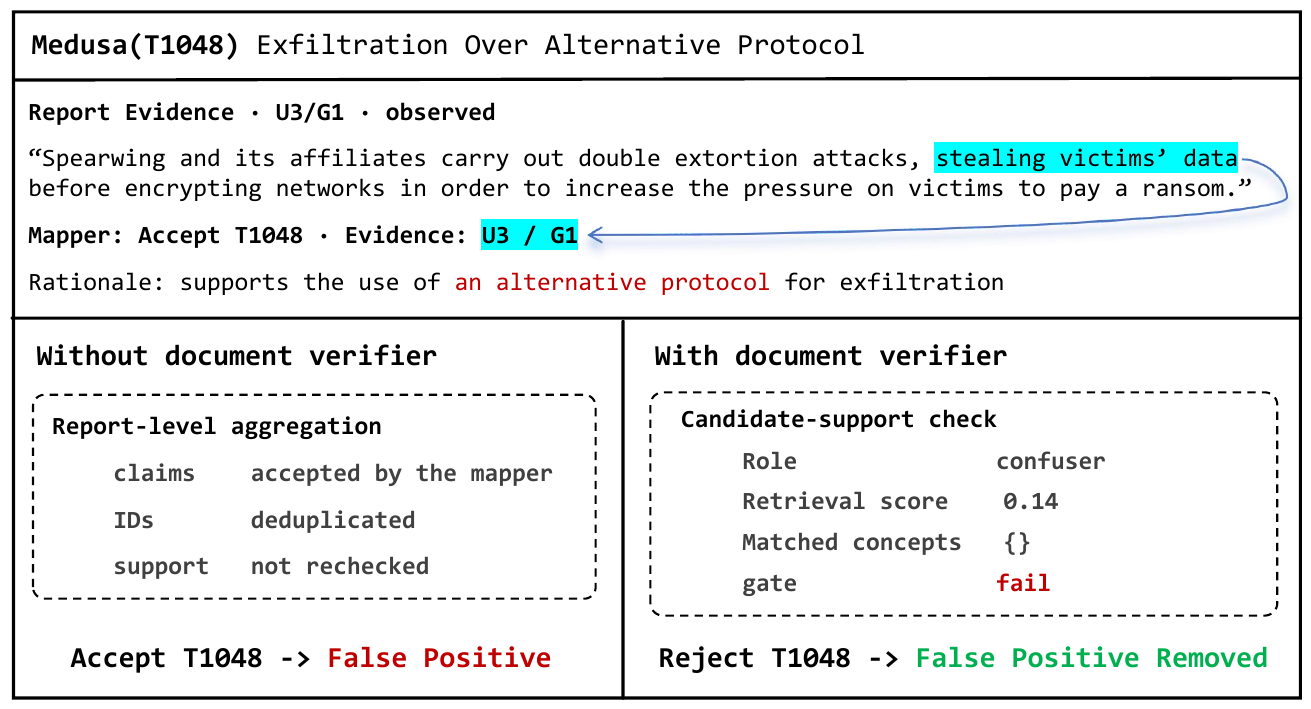}
    \Description{The same Medusa mapper response is compared with and without document-level verification. The verifier rejects T1048 after checking its candidate support.}
    \caption{Document verification for T1048.}
    \label{fig:verifier_case}
\end{figure}

Figure~\ref{fig:verifier_case} illustrates a Medusa claim removed by document verification. Evidence U3/G1 preserves a report statement about data theft before encryption, but the statement does not identify the transfer protocol. The mapper nevertheless accepts T1048 (Exfiltration Over Alternative Protocol), citing U3/G1 and reporting no proof gap. Without the verifier, this assignment remains in the output. The document verifier rejects it because the candidate was retrieved only as an alternative, has a low retrieval score, and has no matched profile concepts. This removes one exact-ID false positive while leaving the evidence, candidates, and mapping response unchanged.

\section{Potential Analyst Workflows}
\label{sec:analyst_workflows}

The structured output of \sys{} supports analyst review beyond a flat label set. For knowledge-base curation, each accepted or abstained technique can be exported with its source URL, quoted spans, evidence status, matched mechanism, and proof gaps. Accepted claims form a review queue; abstentions identify the additional evidence needed. Because the report remains the factual authority, an analyst can reject a mapping without discarding the extracted observation.
The same source-linked evidence can provide bounded hypotheses for code or binary analysis---for example, directing attention from a persistence claim to registry-write routines and startup paths. Future studies could assess these downstream workflows through blinded expert review of independent reports, focusing on claim correctness and analyst review time. Evaluating the code-analysis workflow would additionally require paired reports and malware samples.

\section{Additional Related Work}
\label{sec:additional_related_work}
\textbf{Threat Behavior and Knowledge-Graph Extraction.}
iACE analyzes grammatical relations in technical articles to extract indicators of compromise together with their contextual information~\cite{liao2016acing}. EXTRACTOR simplifies report text, resolves references such as pronouns, and constructs provenance graphs connecting system entities through reported actions~\cite{satvat2021extractor}.
ThreatRaptor constructs behavior graphs from extracted indicators and their relations, then synthesizes queries over system audit logs for threat hunting~\cite{gao2021enabling}. AttacKG aggregates information across reports to enrich technique templates with different behavioral variants~\cite{li2022attackg}.

\noindent\textbf{LLMs for CTI Extraction and ATT\&CK Mapping.}
CTINexus groups entities by type, merges equivalent mentions, and predicts missing relations between subgraphs~\cite{cheng2024ctinexus}. CTIKG normalizes triples and uses a checker agent to identify common output errors and request corrections~\cite{huang2024ctikg}. Fieblinger et al. compare prompting and fine-tuning for CTI triple extraction, using the extracted knowledge for graph construction and link prediction~\cite{fieblinger2024actionable}.
Fayyazi et al. compare encoder-model fine-tuning with decoder-model RAG for identifying the intended tactics of cyberattack procedures~\cite{fayyazi2024advancing}. Guru et al. study LLM-based TTP extraction and embedding-based technique matching for threat-actor attribution~\cite{guru2025technique}.

\noindent\textbf{Evidence-Supported Generation and Verification.}
ALCE evaluates long-form answers with citations, measuring fluency, correctness, and citation quality~\cite{gao2023citations}. Its citation recall measures whether cited passages jointly support a statement, while citation precision identifies irrelevant citations. It uses natural language inference to assess citation support.
RARR checks evidence support after generation~\cite{gao2023rarr}. It constructs search queries from the output, retrieves evidence, and revises unsupported statements while preserving the rest of the text as much as possible.

\end{document}